# GRAVITATIONAL INTERSTELLAR SCINTILLATION

**Redouane Al Fakir**
*The MISC Institute, Vancouver, BC, Canada*
*and*
*Gravitation & Cosmology Group, Dept. of Physics & Astronomy*
*University of British Columbia, Vancouver, BC, Canada*

**Gravitation could modulate the interstellar scintillation of pulsars in a way that is analogous to refractive interstellar scintillation (RISS). While RISS occurs when a large ionized cloud crosses the pulsar line-of-sight, gravitational interstellar scintillation (GISS) occurs when a compact gravitational deflector lies very near to that line-of-sight. GISS differs from RISS in at least two important respects: it has a very distinctive and highly predictable time signature, and it is non-dispersive. We find two very different astronomical contexts where GISS could cause observable diffraction pattern distortions: (1) Highly inclined binary pulsars; GISS-induced displacements up to $10^4 \, km$. Here, pulsar intensity modulations are due to gravitational deflections by the neutron star companion. The double pulsar system PSR J0737-3039 has the ideal configuration for testing for this type of GISS. In addition, because in that remarkable system both neutron stars are pulsars, they can take turns playing the role of scintillating pulsar and gravitational modulator of that scintillation. (2) Extreme scattering events (ESE's); GISS-induced displacements up to $10^6 \, km$. Here, the deflectors are the compact interstellar clouds thought to be responsible for pulsar ESE's. We show that the purported AU size and Jovian mass of these clouds conspire nontrivially to make them good GISS candidates. We conclude that the non-dispersive, dynamically deterministic GISS may complement RISS as a investigative tool when deflectors are present that are either extremely compact, or just compact enough to be significantly gravitating.**

The modulation of the interstellar scintillation of a pulsar by a gravitational source has been proposed before as a means to directly detect gravitational waves. It was concluded that, through the interstellar medium, the exceedingly faint gravitational waves might in a way manifest themselves macroscopically [1].

Unlike the present letter, [1] did not consider the full gravitational field of the deflector, but only its radiative fluctuations. To be potentially observable, the pulsar and the foreground gravity wave source needed to be aligned extremely closely. Some existing chance conjunctions were suggested, and certainly much better ones would be revealed by a targeted search, given the rapidly increasingly number of discovered pulsars. But the deflection-based gravity wave detection schemes in [1,2,3] could not exploit one of the best known cases of natural alignment: the case where a pulsar and a gravitational source are locked into a tight, almost eclipsing multiple star. This class of systems, and specifically tight-orbit, almost eclipsing binary pulsars, were first identified in [4] as ideal for the detection of neutron-star gravity waves through pulsar timing. The specific wishes in [4] for such a system seemed to be answered spectacularly [5] by the discovery of the double pulsar PSR J0737-3039 [6].

In the present letter, we point to a different gravitational deflection-based phenomenon that this time may well be detectable in systems such as PSR J0737-3039. As the deflector, we consider this time the main gravitational field of the pulsar's companion, not its tiny gravity wave fluctuations. The effect might also be detectable in a completely different context, for example if the deflector is a compact interstellar cloud that happens to cross the pulsar line-of-sight (see below).

The gravitational modulation of interstellar scintillation that we suggest here is closely analogous to refractive interstellar scintillation (RISS) [7]. Some important differences include the fact that this gravitational effect is not dispersive, and that it often has a distinctive time dependence that can be determined in advance.

Like its refractive counterpart, though, the present gravitational interstellar scintillation (GISS) starts with the ordinary interstellar scintillation of a pulsar: The pulsar radio waves are scattered by interstellar electron density inhomogeneities of size $a$. If the distance to the Earth is $L > a/\theta_s$ where $\theta_s$ is the scattering angle, then the Earth would be bathing in a diffraction pattern of intensity maxima and minima of typical size $S$. As the Earth moves through the pattern with a relative velocity $V$, the average pulse intensity appears to fluctuate on a timescale $t_S = S/V$. That is ordinary (diffractive) interstellar scintillation (See [8,9] and references therein.)

However, if an inhomogeneity is large enough that $L < a/\theta_s$, then it does not contribute to the diffractive scintillation. Rather, it causes an effective refraction $\theta_r$ of the ray bundles, and ultimately a large-scale displacement of the diffraction pattern (the pattern due to smaller inhomogeneities.) As the Earth moves with relative velocity $V$ through the diffractive pattern, $\theta_r$ varies slowly, inducing a very large-scale refractive focusing of the pattern. This focusing translates in turn into deep modulations of the pulsar intensity over timescales reaching weeks or even months. This is the familiar RISS, which has been extensively confirmed by observation [7-10].

GISS occurs if, instead of a large (refracting) interstellar cloud, one has a foreground compact gravitational source lying very close to the pulsar line-of-sight. Large-scale modulations in pulsar intensity may be caused this time by the spatial and temporal variations of a gravitational deflection angle $\theta_g$, combined with the Earth's velocity $V$ relative to the pulsar diffraction pattern. As in [1], this observer-based deflection angle is related to the Einstein gravitational deflection $\alpha_g$ by the lens equation

(1) $\quad \theta_g \approx \dfrac{D_{GP}}{D_{GE}} \alpha_g$ ,

where $D_{GP}$ is the distance of the gravity source to the pulsar, and $D_{GE}$ is its distance to the Earth.

When the only deflectors along the pulsar line-of-sight are the relatively small ionised clouds responsible for diffractive scintillation (typical size $a < L\theta_s$), the electromagnetic field can be written as [10]

(3) $\quad E(x) = I_0^{1/2} \int dx' \, K_0(x-x') \, e^{i\phi_d(x')}$ .

$I_0$ is the intensity of incident plane waves. $\phi_d$ is the diffractive phase. $x$ is the spatial coordinate of an intensity peak on a one-dimensional screen. The propagator can be written as a function of the Fresnel radius $r_{F0} = (L\lambda)^{1/2}$, where $\lambda$ is the electromagnetic frequency, as

(4) $\quad K_0(x) = \dfrac{\exp[i\pi(x/r_{F0})^2]}{r_{F0}}$ .

In the presence of a gravitational deflector along the line-of-sight, the phase shifts by a small amount $\phi_g$ that can be expanded about a reference position $x_0$:

(5) $\quad \phi_g(x) = \phi_0(x_0) + \phi_1(x-x_0) + \phi_2(x-x_0)^2 + ...$,

The pulsar apparent angular position shifts by

(6) $\quad \theta_g \approx (x_g - x)/D_{GE} = \delta x_g / D_{GE}$,

where $x_g$ is the new position of the intensity peak:

(7) $\quad x_g = G(x - x_0) + x_0 - D_{GE}\lambda\phi_1$

The resulting intensity gain is $G = (1 + D_{GE}\lambda\phi_2)^{-1}$, and the Fresnel radius becomes $r_F = (GL\lambda)^{1/2}$. The electromagnetic amplitude becomes

(8) $\quad E_g(x) = (GI_0)^{1/2} e^{i\psi} \int dx' K_g(x_g - x') e^{i\phi_d(x')}$.

$\psi$ is a new overall phase and the propagator is now

(9) $\quad K_g(x) = \dfrac{\exp[i\pi(x/r_F)^2]}{r_F}$.

The lowest order effect (i.e. considering $G \approx 1$) is then the quasi-rigid displacement of the diffraction pattern:

(10) $\quad E_g(x + \theta_g D_{GE}) \approx E(x)$,

Thus, the gravitational source superimposes a steering effect on the pre-established diffraction pattern that causes it to move quasi-rigidly by distances of order (see (1))

(11) $\quad \delta x_g \approx \theta_g D_{GE} \approx \alpha_g D_{GP}$.

Observationally, the effect is of more interest from an interstellar scintillation physics point-of-view if $\delta x_g > S$, where $S$ is the typical size of features in the pulsar diffraction pattern. If $\tilde{\lambda}$ is the electromagnetic frequency of observation, then

(12) $\quad S = \dfrac{\tilde{\lambda}}{2\pi\theta_S} \propto \dfrac{1}{\tilde{\lambda}}$,

At, say, $\tilde{\lambda} \sim 10$ m, observations would be expected to yield typically $\theta_S \sim 0.2"$ and

(13) $\quad S \sim 300$ km.

CASE 1: HIGHLY INCLINED BINARY PULSARS (PSR J0737-3039)

Consider a very compact gravitational source such as a neutron star, and consider that a companion pulsar is orbiting that gravitational source in a plane that is highly inclined with respect to the Earth. Then, the electromagnetic rays from the pulsar will graze the neutron star at a very small impact parameter *b*, and thus experience a relatively strong deflection (by gravitational standards) of magnitude

(14) $\quad \alpha_g \approx \dfrac{GM}{b}(1+\gamma)(1+\cos\sigma).$

*M* is the mass of the gravitational deflector. *G* is the gravitational constant. $\gamma$ is a parameter that equals 1 if general relativity describes exactly the action of gravity. $\sigma$ is the angular separation between the pulsar and the gravitational source, as seen from the Earth. Also, for simplicity, we have left out an additional deflection component due to the angular momentum of the gravitational source. That component can become significant when the deflector is a rapidly rotating neutron-star.

In our case, the deflector is close to the pulsar line-of-sight, and we assume general relativity. Therefore, we can write **(14)** in the convenient form

(15) $\quad \alpha_g \approx \dfrac{(M/M_\Theta)}{(b/R_\Theta)} 1.75 \ arcsec,$

where $M_\Theta$ and $R_\Theta$ are the solar mass and radius.

In the case of the double pulsar PSR J0737-3039, the binary is almost edge-on, and both neutron stars are pulsars, therefore one has a unique situation in the sense of GISS, where the two neutron stars take turns being the gravitational source and the source of the deflected pulses. Furthermore, the impact parameter is considerably smaller than a solar radius over a substantial part of the orbit, leading to possibly observable effects through **(11)** and **(15)**.

To obtain precise GISS predictions for PSR J0737-3039, several further considerations need to be taken into account, including the precise geometry of the system relative to the Earth at various phases in the orbit and the very different nature of the magnetospheres around the two pulsars. But let us first obtain an order-of-magnitude estimate, and determine if the parameters fall within an observationally interesting range. That is to say, let us see if the binary pulsar can exhibit GISS distortions that are comparable or hopefully larger than a typical pulsar diffractive pattern scale *S* of, say, 300 *km* (see above).

First, consider Pulsar B as the deflector. Its mass is $M \approx 1.25 M_\Theta$. Because of its extended magnetosphere, the smallest impact parameter of Pulsar A must be $b\sim 10{,}000$ *km*. Then, **(15)** would yield $\alpha_g \approx 150 \ arcsec$. On the other hand, the separation between the two pulsars happens to be $D_{GP} \sim R_\Theta$, so that **(11)** yields

(16) $\quad \delta x_g \sim 500 \ km.$

This GISS displacement seems to fall within the range of typical scintillation pattern spatial scales, even tough it was obtained from a combination of parameters that are unrelated to diffractive scintillation. (For example, diffractive spatial scales such as **(13)** depend on the wavelength of observation, while **(16)** does not.) This is a priori a good sign for the observational potential of GISS.

The numbers are better still if we consider the other half of the orbital period, when Pulsar A plays the role of deflector. Its magnetosphere is about 100 times smaller than that of Pulsar B. Its mass ($M \approx 1.34 M_\odot$) is about the same as Pulsar B, but it lets through pulses down to a much smaller impact parameter, in the range $b \sim 500\ km$. The order-of-magnitude estimate for the GISS in this case is a perhaps remarkable

**(17)**    $\delta x_g \sim 10,000\ km$.

This is of course the maximal amount of GISS-induced spatial displacements in the diffraction pattern. What is usually targeted by observation is the focusing (and hence the modulation in the amplitude of pulsar intensity peaks) that results from differentials in $\delta x_g$ as it oscillates between **(17)** and much smaller values, as a result of the complicated (but predetermined) time variations of $\alpha_g$ over one orbital period.

CASE 2: COMPACT CLOUDS IN PULSAR EXTREME SCATTERING EVENTS

Twenty years after the discovery of the first extreme scattering events (ESE) of compact radio sources [11], their origin and properties remain uncertain. So far, the most widely discussed scenario is that an ESE happens when a supercompact interstellar cloud of perhaps $M \sim 10^{-3} M_\odot$ and size $R \sim 1\ AU$ crosses the line-of-sight of a pulsar or a quasar [12-15].

Because of the point-like size of pulsars, their ESE's can be investigated with the powerful tool of interstellar scintillation [12-20]. Let us see if the numbers are right for GISS to be potentially at play in this important class of phenomena. As above, the test is whether the cloud parameters produce GISS diffraction pattern displacements in **(11)** that are larger (hopefully much larger) than the pattern's spatial scale **(12)**.

Using the typical parameter values above in **(11)** and **(15)**, we obtain the formula for maximal diffraction pattern displacements caused by GISS in the context of ESE's:

**(18)**    $\delta x_g \approx \dfrac{M}{10^{-3} M_\odot} \dfrac{1\ AU}{b} \dfrac{D_{GP}}{1\ Kpc} \times 1.2 \times 10^6\ km$.

Thus, perhaps surprisingly, GISS can produce displacements as large as $\delta x_g \sim 10^6\ km$, which is a range more familiar in regular RISS. (Surprising, because gravitational deflection is usually assumed to be much smaller than dispersive refraction in interstellar cloud physics.)

An important consequence of **(18)** is that GISS may be significant even at large impact parameters ($\delta x_g \sim 10^4 \, km$ at $b \sim 100 \, AU$). Therefore, if the compact cloud scenario is indeed the right explanation for ESE's, then these events should be preceded *and* followed by a GISS episode.

Thus, GISS might provide an additional tool for investigating these puzzling ESE's, especially through the prediction of a non-dispersive, distinctive time signature in the pulsar scintillation during ingress and egress.

More generally, GISS may complement RISS as an investigative tool when interstellar deflectors are too compact, be they stars or interstellar clouds. One distinctive feature of GISS (besides being non-dispersive and more deterministic than RISS), is that it kicks in well before the pulsar line-of-sight actually crosses the interstellar deflector responsible for the scintillation event, and lasts well after.